\documentclass[twoside]{ae100prg}
\usepackage{amsmath}
\usepackage{bm}
\usepackage{graphicx}
\usepackage[breaklinks]{hyperref}
\usepackage{booktabs}

\begin{document}
\title{Illusory horizons, thermodynamics, and holography inside black holes}


\author{Andrew J. S. Hamilton}

\address{JILA, Box 440, U.~Colorado, Boulder, CO 80309, USA}

\email{Andrew.Hamilton@colorado.edu}

\newcommand{\dd}{d}
\newcommand{\ee}{e}
\newcommand{\im}{i}

\newcommand{\emit}{{\rm em}}
\newcommand{\inn}{{\rm in}}
\newcommand{\obs}{{\rm obs}}
\newcommand{\out}{{\rm ou}}
\newcommand{\rem}{r_\emit}
\newcommand{\robs}{r_\obs}

\newcommand{\bx}{\bm{x}}
\newcommand{\by}{\bm{y}}

\hyphenpenalty=3000

\newcommand{\penrosefig}{
    \begin{figure}[bt!]
    \centering
    \includegraphics[scale=1]{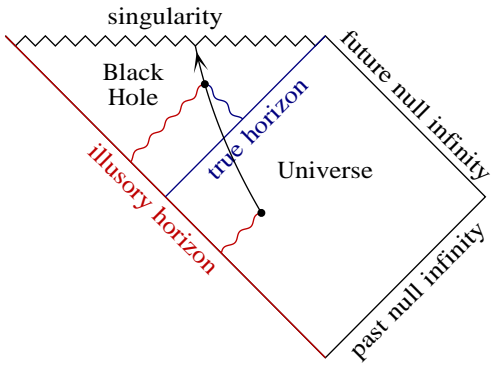}
    \caption[1]{
    \label{penrose}
Penrose diagram
of a Schwarzschild black hole.
The arrowed line represents the trajectory of an observer,
while the wiggly lines represent light rays
perceived by the observer
from the illusory (red) and true (blue) horizons.
    }
    \end{figure}
}

\newcommand{\schwfig}{
    \begin{figure}[btp!]
    \centering
    \includegraphics[width=2.25in]{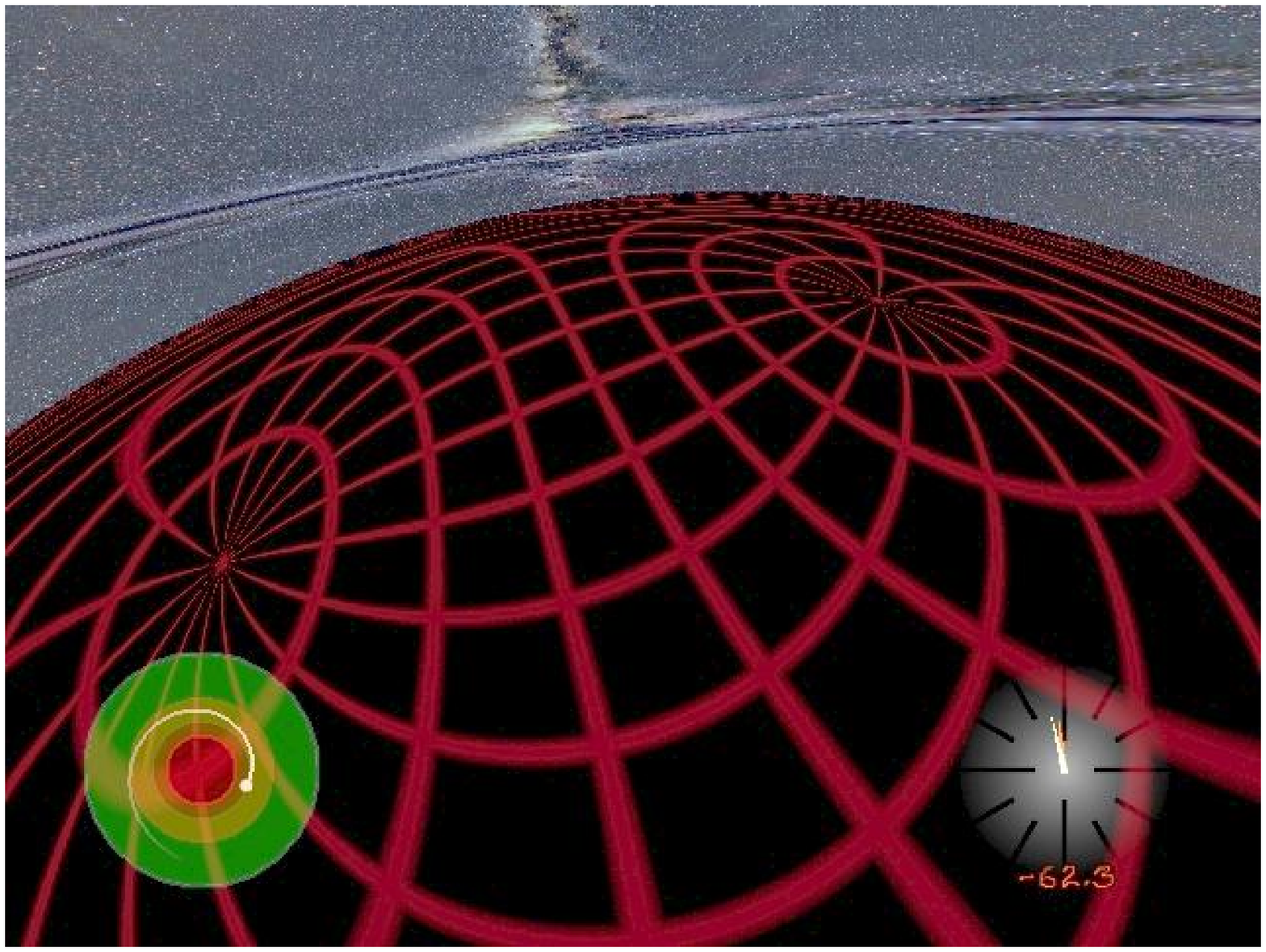}
    \includegraphics[width=2.25in]{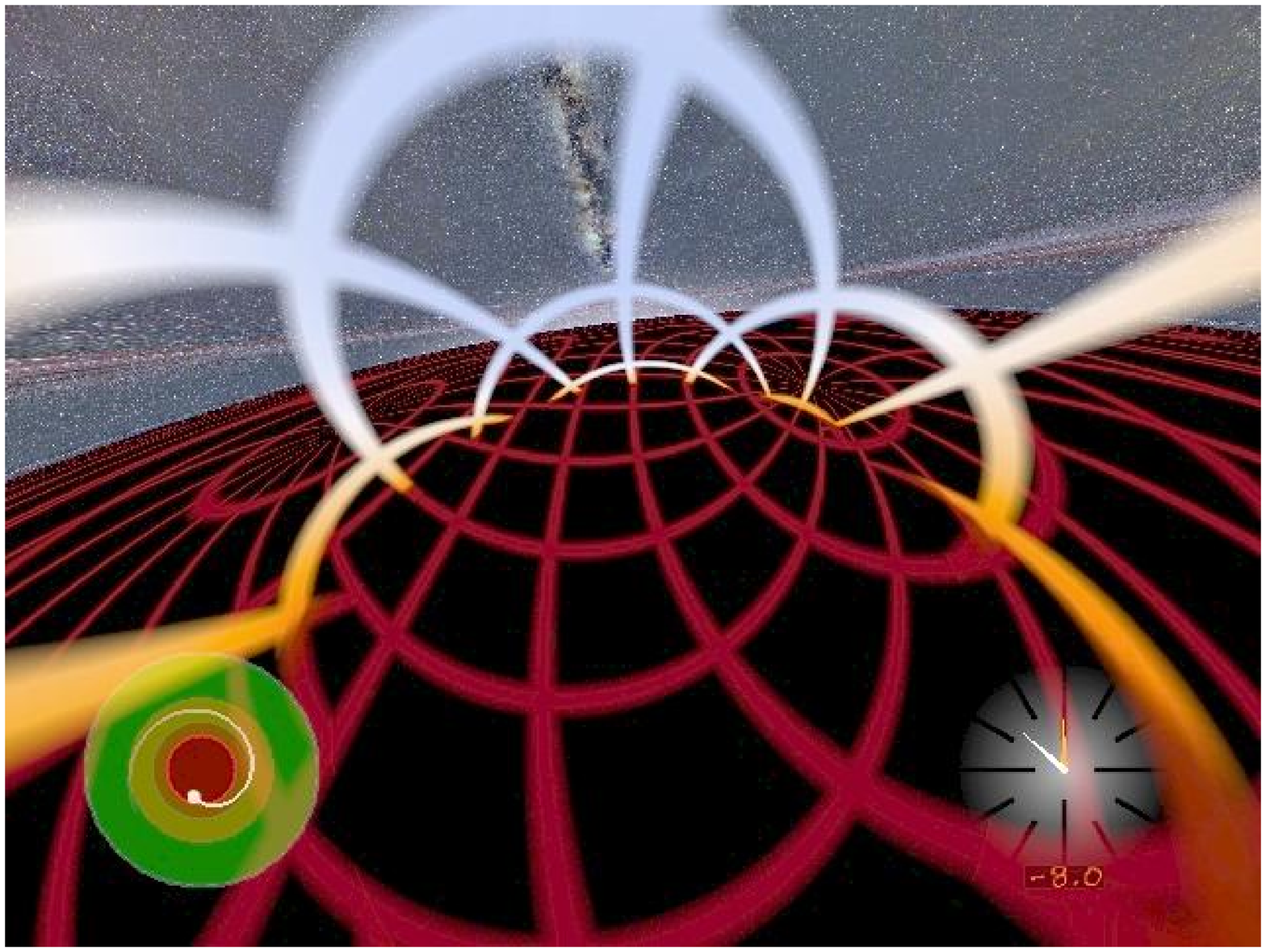}
    \includegraphics[width=2.25in]{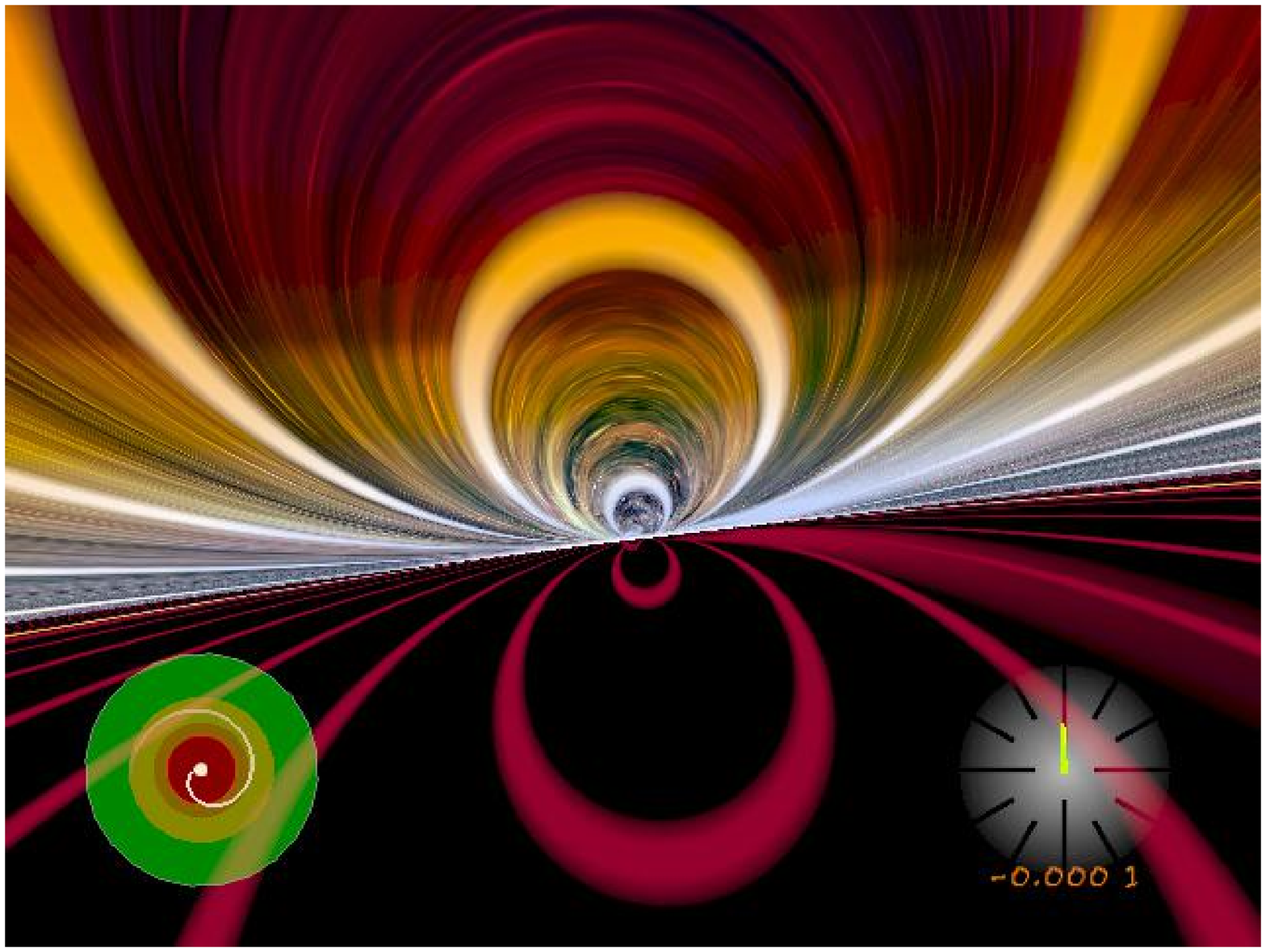}
    \caption[1]{
    \label{schw}
Visualization of the scene seen
by an observer falling into
a Schwarzschild black hole
on a geodesic with specific energy and angular momentum
$E = 1$ and $L = 3.92$ geometric units,
from \cite{Hamilton:2010my}.
In the upper panel,
the observer is at a radius of
$3.000$,
outside the true horizon;
in the middle panel
the observer is at a radius of
$1.613$,
inside the true horizon;
in the bottom panel
the observer is at a radius of
$0.045$,
near the central singularity.
The illusory horizon is painted with a dark red grid,
as befits its infinitely redshifted appearance,
while the true horizon is painted with an
appropriately red- or blue-shifted blackbody color.
Further frames and details of this visualization are at
\cite{Hamilton:2010my}.
The background is Axel Mellinger's Milky Way
\cite{Mellinger:2009asp}
(with permission).
    }
    \end{figure}
}

\newcommand{\collfig}{
    \begin{figure}[tb!]
    \begin{center}
    \leavevmode
    \includegraphics[bb=91 183 519 610,scale=.31]{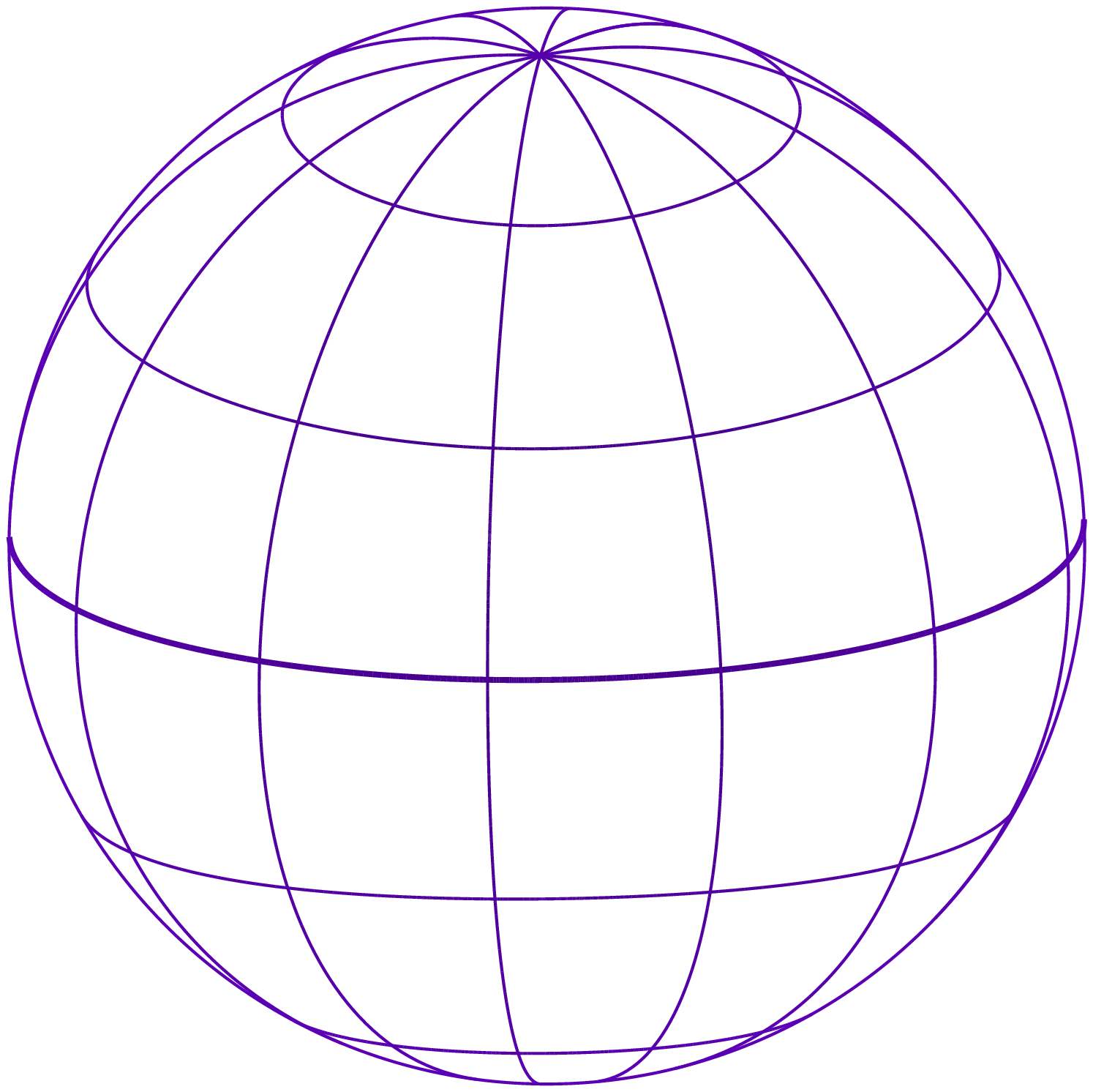}
    \hspace{1mm}
    \includegraphics[bb=196 183 415 610,scale=.31]{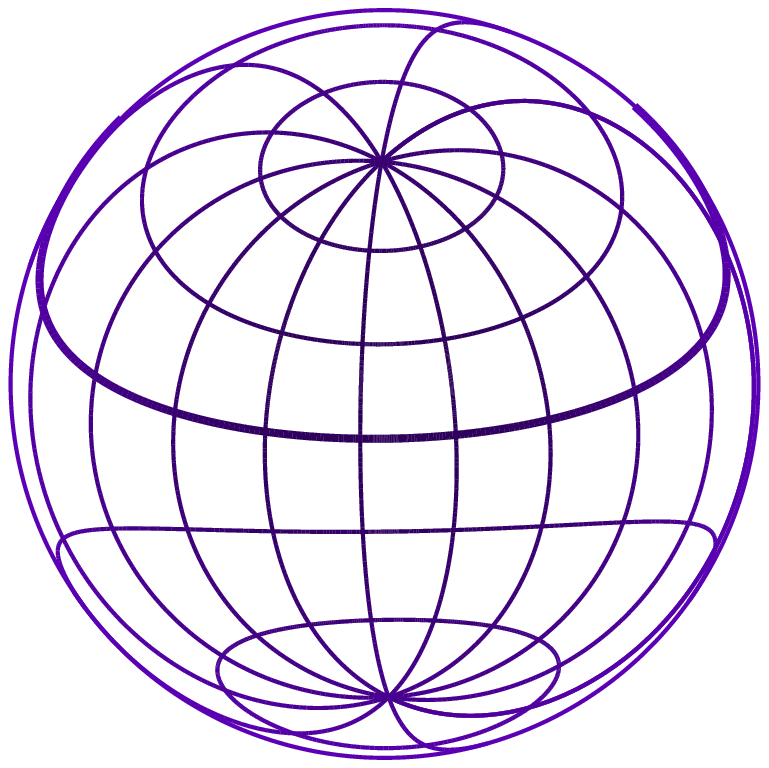}
    \hspace{1mm}
    \includegraphics[bb=223 183 521 610,scale=.31]{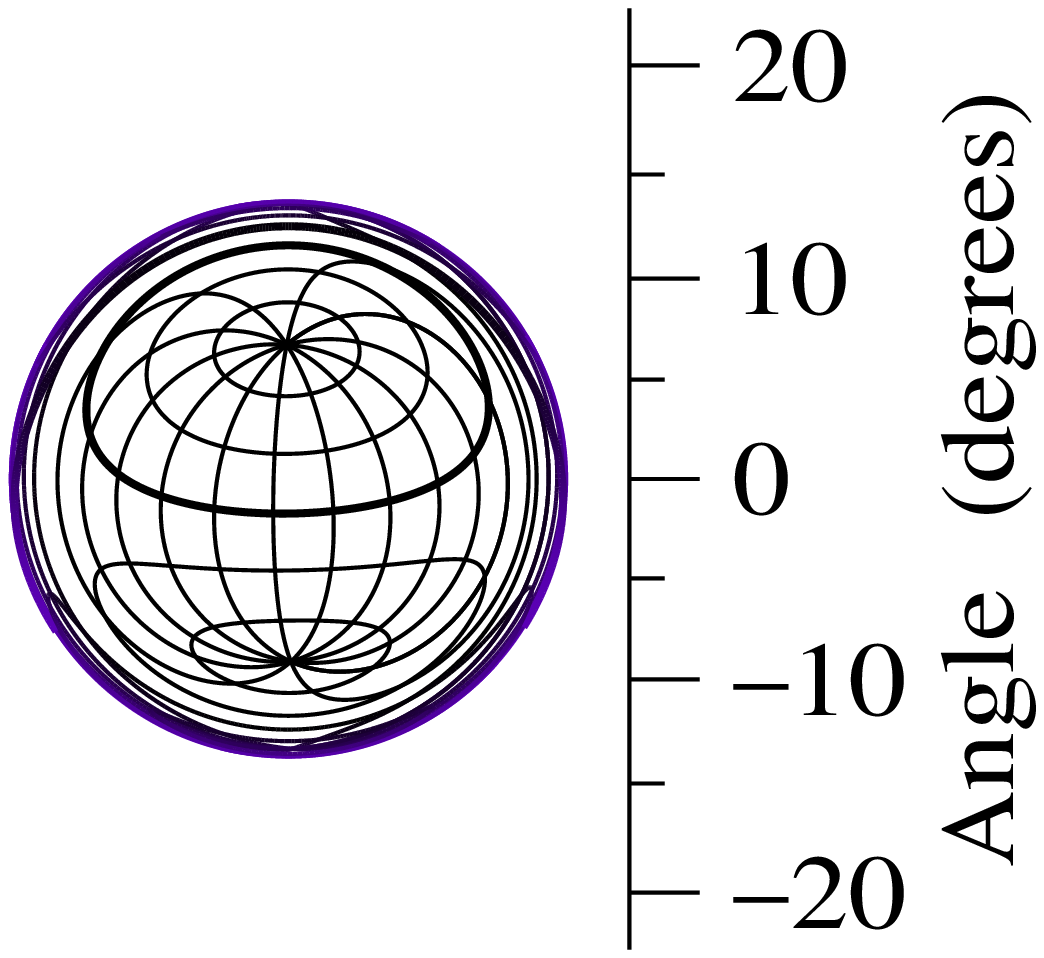}
    \end{center}
    \caption[Collapse to a black hole]{
    \label{coll}
Three frames in the Oppenheimer-Snyder
collapse of
\cite{Oppenheimer:1939ue}.
a star,
as seen by an outside observer at rest
at a radius of $20$ geometric units.
As time goes by, from
left to right,
the collapsing star appears to freeze at its horizon,
and take on the appearance of a Schwarzschild black hole.
An animated version of this visualization is at
\cite{Hamilton:1998coll}.
    }
    \end{figure}
}

\newcommand{\kappazerofig}{
    \begin{figure}[bt!]
    \centering
    \includegraphics[scale=.6]{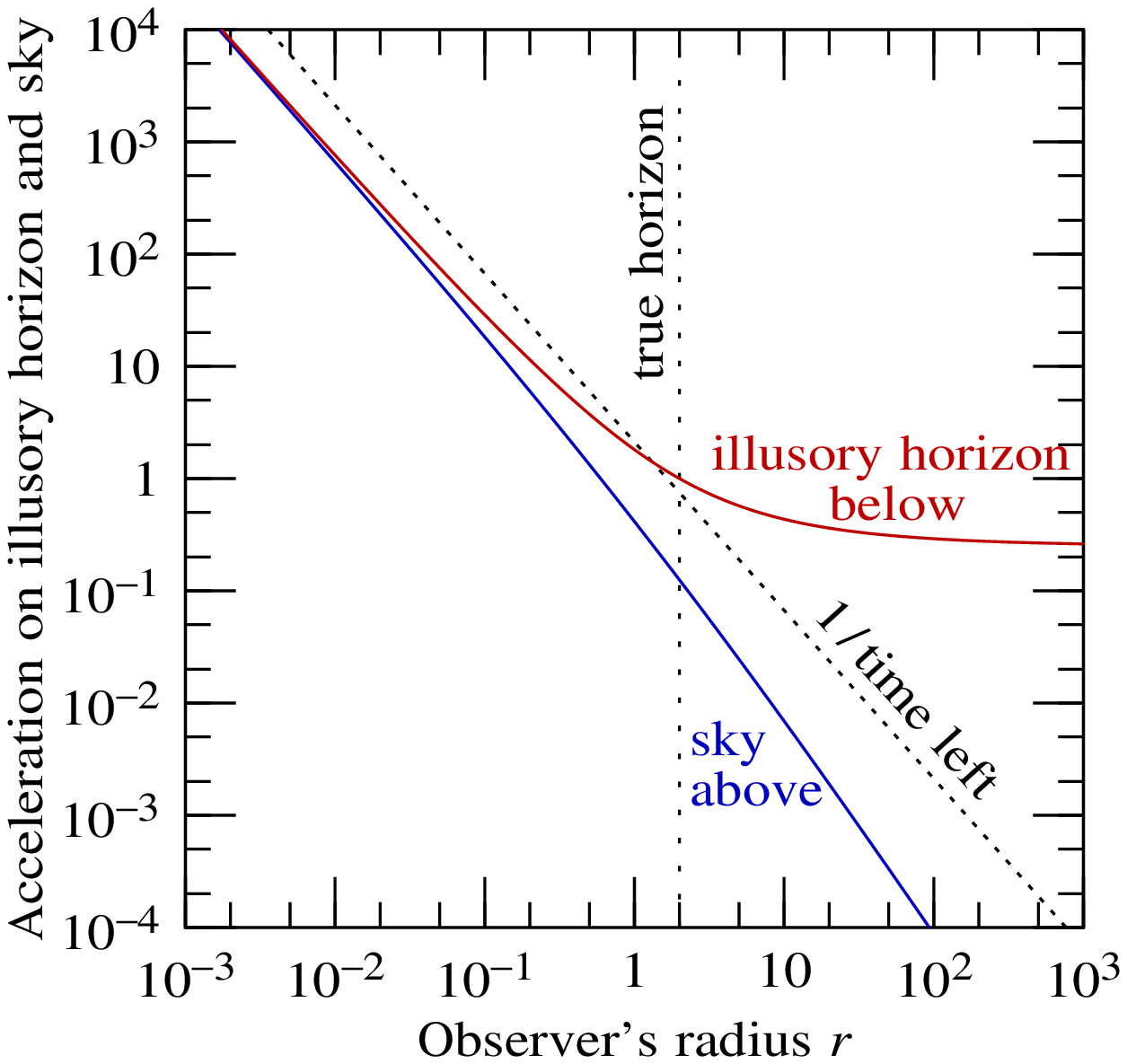}
    \caption[]{
    \label{kappa0}
Acceleration at the illusory horizon
directly below,
and at infinity directly above,
seen by a radially free-falling infaller at radius $r$.
The dashed line shows the reciprocal of the proper time
left until the infaller hits the singularity.
The acceleration diverges towards the singularity $r \rightarrow 0$,
suggesting a logarithmic divergence in the total number of
Hawking quanta observed by an infaller reaching the singularity.
    }
    \end{figure}
}

\newcommand{\kappahfig}{
    \begin{figure}[bt!]
    \centering
    \includegraphics[scale=.6]{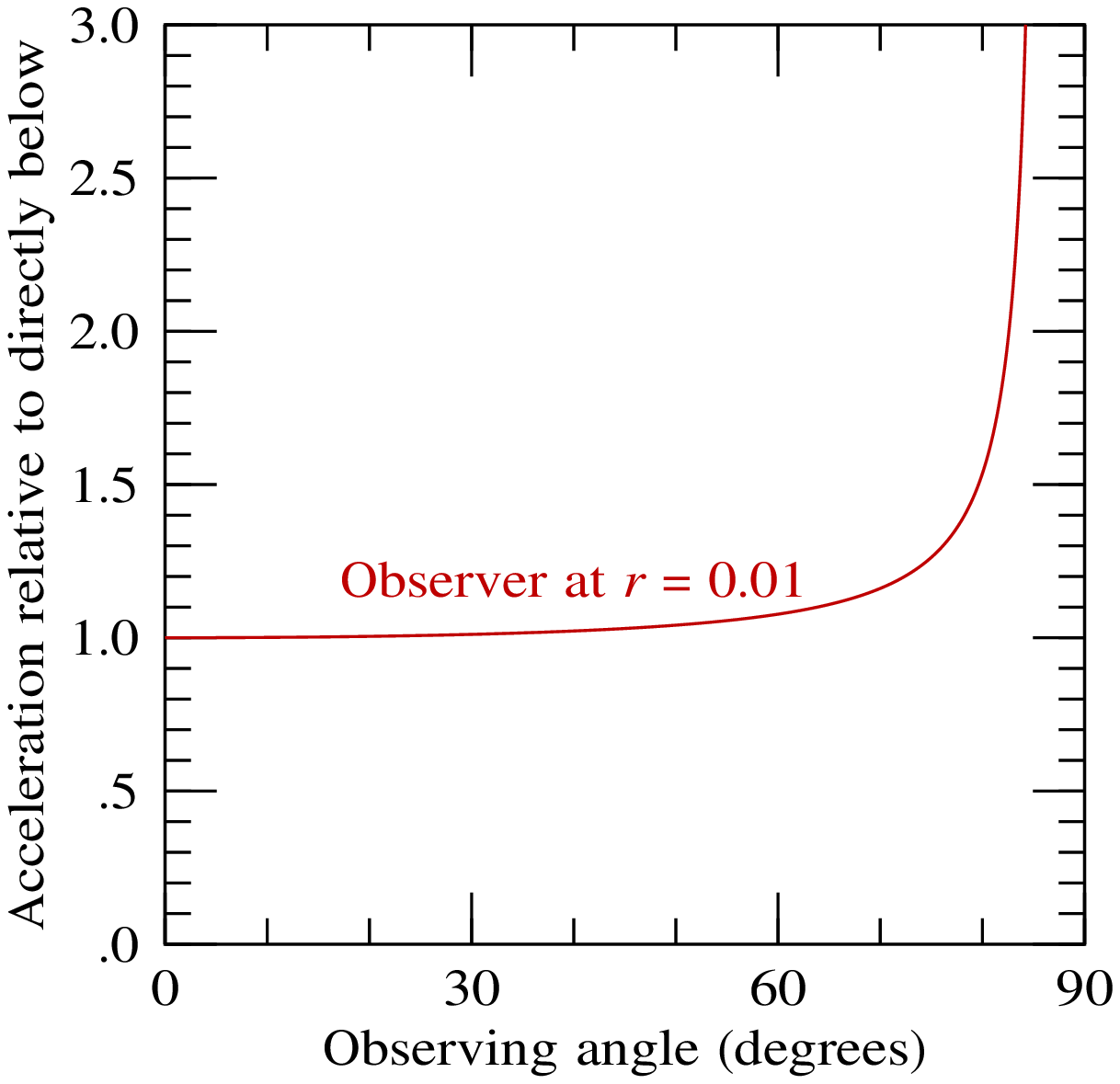}
    \caption[]{
    \label{kappah}
Acceleration $\kappa$ on the illusory horizon
seen by a radially free-falling non-rotating infaller,
relative to the acceleration $\kappa_0$ directly below
(towards the black hole),
as a function of the viewing angle
relative to directly below.
The example curve shown is as seen by an infaller well inside the horizon,
at radius $0.01$ geometric units.
The acceleration is constant out to near
the perceived edge of the black hole,
where the acceleration diverges.
Curves at other radii are similar.
    }
    \end{figure}
}

\newcommand{\bhentropyfig}{
    \begin{figure}[bt!]
    \centering
    \includegraphics[scale=.9]{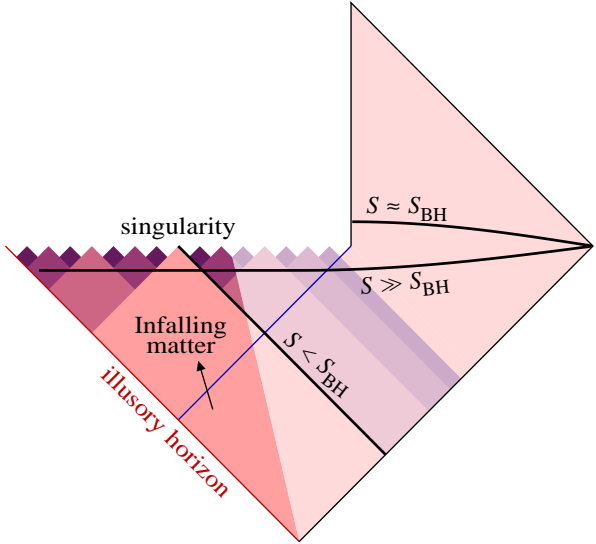}
    \caption[Locality implies a violation of the second law inside black holes]{
    \label{bhentropy}
Near its singularity,
a black hole contains numerous regions whose future lightcones do not intersect.
If locality held inside a black hole,
then it would be legimitate to accumulate entropy along a spacelike surface
slicing through these causally disconnected regions.
Dissipative processes inside a black hole
can potentially cause the entropy accumulated along the spacelike surface
to exceed greatly the Bekenstein-Hawking entropy of the black hole
\cite{Wallace:2008zz},
leading to a violation of the second law when the black hole evaporates.
This argument strongly supports the idea that locality
must break down inside black holes.
Whereas entropy passing through a spacelike surface inside the black hole
may exceed the Bekenstein-Hawking entropy,
the entropy passing through any null surface inside the black hole
is always less than the Bekenstein-Hawking entropy,
consistent with Bousso's
\cite{Bousso:2002ju}
covariant entropy bound.
    }
    \end{figure}
}

\newcommand{\pairfig}{
    \begin{figure}[bt!]
    \centering
    \includegraphics[scale=.9]{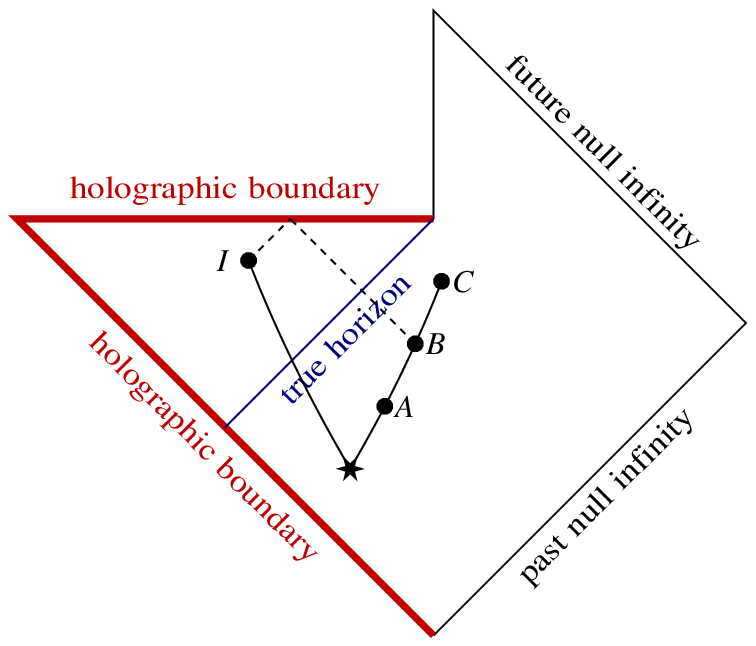}
    \caption[Penrose diagram of entangled pair]{
    \label{pair}
The illusory horizon and the singularity
constitute the holographic boundary of an evaporating black hole.
The diagram illustrates
the delocalization of an entangled pair
created at the star point.
Locality holds between an inside observer $I$
and an outside observer
as long as their future lightcones intersect,
so that they can communicate before $I$ hits the singularity.
Thus locality holds between $I$ and $A$,
is at the brink of failure between $I$ and $B$,
and fails between $I$ and $C$.
    }
    \end{figure}
}

\begin{abstract}
There is persistent and endemic confusion
between the true (future) horizon
and the illusory (past) horizon of a black hole.
The illusory horizon is the redshifting surface of matter
that fell into the black hole long ago.
A person who free-falls through the horizon of a black hole falls
through the true horizon,
not the illusory horizon.
The infaller continues to see the illusory horizon ahead of them,
all the way down to the classical singularity.
The illusory horizon is the source of Hawking radiation,
for both outsiders and infallers.
The entropy of a black hole is $1/4$ of the area of the illusory horizon,
for both outsiders and infallers.
The illusory horizon holographically encodes states hidden behind it,
for both outsiders and infallers.
The endpoint of an infaller approaching the classical singularity
is to merge their states with the illusory horizon.
The holographic boundary of the black hole is then the
union of the illusory horizon and the classical spacelike singularity.
When an infaller reaches the classical singularity,
any entanglement of the infaller with outsiders or other infallers
is transferred to entanglement with the states of the black hole,
encoded on the illusory horizon.
Locality holds between an infaller and a spacelike-separated
outsider or other infaller as long as their future lightcones
intersect before the singularity,
but breaks down when the future lightcones no longer intersect.
\end{abstract}

\section{Introduction}
\label{intro-sec}

There is persistent and endemic confusion
in the literature between the true (future) horizon
and the illusory (past) horizon of a black hole.
The confusion has led to the misconception that
Hawking radiation is emitted from the true horizon,
and that the states of a black hole are encoded on the true horizon.

The presence of a black hole introduces a bifurcation boundary to spacetime,
separating the spacetime into
a region that an observer can see,
and a region that is invisible to the observer.
This bifurcation horizon is the illusory horizon,
and it is observer-dependent.
The illusory horizon is the boundary of the past lightcone of an observer
watching the black hole.

When an observer measures thermodynamic variables such
as temperature or entropy,
they must measure degrees of freedom that are actually available to them,
which is to say, degrees of freedom along their past lightcone.
Thus a consistent description of
generalized thermodynamics by an actual observer
must involve the observer's illusory horizon, not the true horizon.

\penrosefig

The purpose of this paper is to set forward a number of proposals
regarding generalized thermodynamics from the perspective
of observers who fall through the true horizon.
The proposals are motivated by the classical appearance of
the illusory horizon seen by an infaller.
The classical appearance suggests that the principles of
generalized thermodynamics and holography extend to infallers
in the simplest and most obvious way.

For simplicity, this paper considers only a spherically symmetric,
uncharged (Schwarzschild) black hole.

\section{The illusory horizon}
\label{illusory-sec}

\schwfig

Figure~\ref{penrose}
shows the familiar Penrose diagram
of a Schwarzschild black hole,
with the illusory (past) and true (future) horizons labelled.
In the analytically extended Schwarzschild geometry,
the illusory horizon is a true horizon,
the horizon of a white hole and parallel universe.
In a real black hole however,
the Schwarzschild past horizon
is replaced by the exponentially dimming and redshifting image
of the star that collapsed to the black hole long ago.

\collfig

As the Penrose diagram of the Schwarzschild black hole shows,
when an observer outside the black hole looks at the black hole,
they are looking at the illusory horizon.
When an observer free-falls through the horizon of the black hole,
they fall through the true horizon,
not the illusory horizon.
The true horizon becomes visible to the observer
only after the observer has passed through it.
The illusory horizon continues to appear ahead of the observer
even after they have passed through the true horizon.

Figure~\ref{schw}
illustrates three frames from a visualization
of the scene seen by an observer who free-falls into
a Schwarzschild black hole
\cite{Hamilton:2010my,Hamilton:2010schw}.
These scenes are general relativistically ray-traced, not artist's impressions.
The illusory and true horizons of the black hole are painted with grids
of latitude and longitude,
so that they can be seen.
The illusory horizon is of course infinitely redshifted
in the Schwarzschild geometry,
but it is nevertheless possible to ray-trace light rays
from an infinitesimal distance off the illusory horizon.

The visualization confirms the expectation from the Penrose diagram.
When the observer falls through the horizon,
they do not fall through the illusory horizon,
which continues to appear a finite distance ahead of the observer.
Instead, the observer falls through a new entity, the true horizon,
which was invisible until the observer passed through it.
At the moment the observer passes through the true horizon,
it forms a line extending down to the illusory horizon.
As the observer falls inward,
the true horizon expands into a bubble over the observer's head.
The circle where the illusory and true horizons intersect expands.

Are visualizations of the Schwarzschild geometry a reliable guide to
visualizations of real spherical black holes?
Yes.
Figure~\ref{coll}
shows three frames from the collapse of a spherical,
uniform density, pressureless star that starts from zero
velocity at infinity, a problem first solved by Oppenheimer and Snyder
\cite{Oppenheimer:1939ue}.
The frames are as seen by an observer at radius $20$ geometric units.
Again, these frames are general relativistically ray-traced,
not artist's impressions.
The frames take into account the differential light travel time
from different parts of the star's surface to the observer.
As the star approaches its horizon,
the star freezes,
and takes on the appearance of a Schwarzschild black hole.

\kappahfig

\section{The illusory horizon is the source of Hawking radiation, for outsiders and insiders}
\label{hawk-sec}

At its most fundamental level,
Hawking
\cite{Hawking:1975b}
or Unruh
\cite{Unruh:1976db,Crispino:2007eb}
radiation
arises when an observer watches an emitter that is accelerating
relative to the observer.
When waves that are pure negative frequency (positive energy)
in the emitter's frame
are propagated to the observer,
the acceleration causes the waves to appear to be a mix
of negative and positive frequencies in the observer's frame.
In particular,
the emitter's vacuum
(``in'' vacuum)
is not the same as the observer's vacuum
(``out'' vacuum).
A classic calculation
(e.g.\ \cite{Visser:2001kq,Padmanabhan:2010})
shows that if the acceleration is approximately constant
over several acceleration timescales,
then the observer will see the emitter's vacuum
as a thermal state with temperature proportional to the acceleration.

An observer watching a black hole sees Hawking radiation
because matter that collapsed to the black hole long ago
appears classically frozen at the illusory horizon,
apparently accelerating away from the observer,
redshifting and dimming into the indefinite future.
When an infaller free-falls through the true horizon,
they do not encounter the redshifting surface at the true horizon.
Rather,
the infaller sees the redshifting surface of the collapsed matter
continue to remain on the illusory horizon ahead of them,
as illustrated by Figure~\ref{schw}.

\kappazerofig

An exact calculation of the Hawking emission seen by
an infaller is difficult,
as illustrated by the efforts of
\cite{Hodgkinson:2012mr}
reported at this conference.
The reason for the difficulty is that,
whereas for a distant observer
only the monopole mode of emission is important,
for an infaller all angular modes contribute.
However,
it is possible to predict the qualitative character
of the Hawking radiation
from a classical calculation of the acceleration at
the illusory horizon, as witnessed by an infaller.

The acceleration,
hence the Hawking or Unruh radiation,
that an infaller sees depends on the state of motion of the infaller.
The simplest case is that of an observer
who free-falls radially from zero velocity at infinity,
and who fixes their gaze in a particular direction
(that is, the infaller's detector is non-rotating).
Figure~\ref{kappah}
shows the acceleration on the illusory horizon seen by such an infaller
well inside the true horizon,
at a radial position $r = 0.01$ geometric units.
Note that the observer here is staring at a fixed angular direction
relative to their own locally inertial frame,
not at a fixed angular position on the black hole.
Figure~\ref{kappah}
shows that the acceleration is approximately constant
out to near the perceived edge of the black hole,
indicating that the acceleration directly below
is representative of the black hole as a whole.

Figure~\ref{kappa0}
shows the acceleration on the illusory horizon directly below,
as seen by the radially free-falling infaller as a function
of their radial position $r$.
The acceleration is approximately constant ($1/4$ geometric units)
far from the black hole, but increases inward,
diverging as the infaller approaches the classical singularity,
$r \rightarrow 0$.
The Figure shows that
the acceleration changes on a timescale comparable
to the proper time left for the infaller to hit the singularity.
Thus the usual connection between acceleration and temperature
(which requires the acceleration to remain approximately constant
over several acceleration times) fails.
Nevertheless,
the calculation does suggest that the Hawking radiation
witnessed by an infaller might diverge
as the infaller approaches the singularity.
The calculation suggests of order one Hawking quantum per time remaining,
or a logarithmically diverging total number of quanta.
Rigorous calculation will be required to test this proposal.

Figure~\ref{kappa0}
also shows the acceleration on the distant sky directly above,
as seen by the radially free-falling infaller.
The acceleration is negligible when the infaller is far from the black hole,
but increases inward.
Interestingly,
the acceleration on the sky above approaches the same diverging
value as that on the illusory horizon below
as the infaller approaches the singularity.
This suggests that the infaller approaching the singularity
might see logarithmically diverging Hawking radiation from all directions.

\section{The entropy of a black hole is $1/4$ the area of the illusory horizon, for outsiders and insiders}

Generalized thermodynamics (e.g.\ \cite{Wald:1999vt}) postulates that
from the perspective of an observer outside the true horizon,
a black hole that has reached near stationarity
should be treated as an object in near thermodynamic equilibrium,
with an entropy equal to $1/4$ of its horizon area in Planck units,
and a temperature equal to $1/(2\pi)$ times the acceleration
at the illusory horizon.

Generalized thermodynamics may reasonably be expected to
hold also for infallers.
For example, it would be quite extraordinary if an infaller witnessed
a violation of the second law of thermodyamics.
As remarked in the Introduction,
an observer must count entropy that is visible to them,
that is, entropy along their past lightcone.
The boundary of the observer's past lightcone towards the black hole
is the illusory horizon.
Generalized thermodynamics teaches that entropy must be
associated with the boundary,
the illusory horizon.

Figure~\ref{schw}
shows that the appearance of the illusory horizon is seamless
for infallers who free-fall through the true horizon.
It is natural therefore to propose that
the entropy of the black hole is $1/4$
the area of the illusory horizon not only for outsiders,
but also for infallers.
Indeed,
if an infaller saw the horizon entropy decrease when they fell inside,
then that would violate the second law.
Conversely if the infaller saw the horizon entropy increase,
then the black hole would appear to the infaller to contain more entropy than
a quarter its horizon area, contradicting the notion
that a stationary black hole is in a thermal condition
of maximum entropy.

The idea that the illusory horizon, not the true horizon,
is the carrier of the hidden states of the black hole
is consistent with the fact that Hawking radiation
originates from the illusory horizon, not the true horizon.

\section{The illusory horizon is a holographic screen, for outsiders and insiders}

The information paradox originated in a seminal paper by Hawking
\cite{Hawking:1976ra}.
The paradox is that one of two revered principles of quantum field theory
must break down in the presence of black hole horizons:
either locality must fail, or else unitarity must fail.
Locality is the proposition
that spacelike-separated field operators must commute.
Locality ensures that no information can be transmitted
between spacelike-separated points,
enforcing causality at the quantum level.
Unitarity is the proposition that dynamics is reversible at the quantum level.
Hawking tacitly assumed that locality holds,
and showed that the Hilbert space of states inside a black hole is then disjoint from
those of an observer to the future of when the black hole has evaporated.
Consequently information is destroyed, violating unitarity.

\bhentropyfig

The most widely accepted resolution of the information paradox is holography,
an idea originally proposed by
t'Hooft \cite{'tHooft:1993gx}
and Susskind \cite{Susskind:1994vu}.
Holography asserts that the quantum states seen by an insider
are seen by an outsider as residing on the horizon of the black hole.
Holography violates locality
because the Hilbert spaces of spacelike-separated regions,
far from being disjoint, are identified with each other.
Information about what happens inside the black hole
is encoded on its horizon, and eventually radiated
to the outside as Hawking radiation, preserving unitarity.
Holography has received impetus from
gauge/gravity dualities that arise in string theory,
whereby a strongly gravitating system
is dual to a conformal gauge theory residing on the boundary
of the system.

Arguments favouring a breakdown of locality
become stark when one considers not just one insider,
but a succession of infallers.
As shown by
\cite{Wallace:2008zz},
if a black hole accretes gas,
increasing its Bekenstein-Hawking entropy by some amount,
then processes of dissipation inside the black hole can potentially
increase the entropy of the gas not merely by the increase in the
Bekenstein-Hawking entropy,
but rather by some fraction of the total Bekenstein-Hawking entropy
of the entire black hole.
If locality held,
then it would be legitimate to accumulate the entropy
from multiple parcels of infalling gas,
leading to a total entropy inside the black hole many orders of
magnitude greater than its Bekenstein-Hawking entropy.
This would imply a gross violation of the second law
when the black hole subsequently evaporated,
as illustrated by Figure~\ref{bhentropy}.
To save the second law of thermodynamics from the
\cite{Wallace:2008zz}
argument,
locality must be abandoned not only across the horizon,
but between a multiple succession of infallers.

Holography produces just the kind of breakdown of locality
that is needed to save the second law of thermodynamics inside black holes.
Just as an outsider must count states hidden behind their illusory horizon as being
holographically encoded on their illusory horizon,
so also an infaller must count states hidden behind their illusory
horizon as being holographically encoded on their illusory horizon.
In this view,
an infaller should not count the entropy production
witnessed by earlier infallers
if that entropy production occurred behind the later infaller's illusory horizon.

\section{An infaller merges states with the illusory horizon at the classical singularity}

\pairfig

The bottom panel of
Figure~\ref{schw}
shows that, as an infaller approaches the classical singularity,
they have the impression of reaching the illusory horizon,
which gives the appearance of a flat plane.
Any quantitative measure of distance to the illusory horizon,
such as the affine distance
(the affine parameter normalized to measure proper distance
in the observer's frame),
or the angular diameter distance
(the distance inferred from the apparent angular separation of
objects a known distance apart,
such as lines of constant latitude and longitude),
indeed goes to zero as the observer approaches the singularity.

In the light of the classical appearance,
it is natural to propose that an infaller who reaches the singularity
merges their states with the illusory horizon.
It has been argued in this paper that prior to the singularity,
the experience of an infaller can be described by general relativity
coupled with a natural extension of generalized thermodynamics.
Such a description must fail at the singularity,
where the tidal force diverges,
and, as argued in \S\ref{hawk-sec}, the Hawking radiation may also diverge.
The proposal is that the description of physics at the singularity
should be replaced by a holographic dual description.
In this picture, as illustrated in
the Penrose diagram in Figure~\ref{pair},
the complete holographic boundary of the black hole consists
of the union of the illusory horizon and the singularity.

\section{Where locality breaks down inside black holes}

The simplest possibility is that the transition from a classical
to a dual holographic description at the singularity is so rapid
as to be effectively instantaneous.
If so, then any quantum entanglement between an infaller
and an outsider or other infallers
will be replaced ``instantly'' by entanglement
with the holographic image of the black hole
when the infaller hits the singularity.

Figure~\ref{pair}
illustrates how locality between a pair of particles created
in an entangled state
(e.g.\ a spin-zero singlet of spin-up and spin-down particles)
breaks down as one of the pair falls inside the black hole
towards the singularity.
Locality holds between an insider who observes the inside particle at $I$,
and an outsider who observes the outside particle at $A$,
because their future lightcones intersect,
so they can compare their measurements of spin.
But locality fails between $I$
and an outsider who observes the outside particle at $C$,
because their future lightcones do not intersect,
so it is too late to compare measurements.
The transition between locality and non-locality takes place at $B$,
where the future lightcones just intersect at the singularity.

\section{Summary}

In this paper I have presented several arguments
and proposals about generalized thermodynamics and holography
from the point of view of observers who fall through the true horizon
of a black hole.
The proposals are motivated by the classical appearance of a black hole
as seen by an infaller.
The proposals are consistent with, and extend,
prevailing popular ideas about generalized thermodynamics and holography
from the point of view of observers who remain outside the horizon.

An important point is that observers see Hawking radiation
not from the true (future) horizon,
but from the illusory (past) horizon,
which is the redshifting surface of matter that fell into the black hole
long ago.
The illusory horizon is the boundary of the past lightcone of an observer,
and is observer-dependent.
The illusory horizon is the holographic screen of the black hole
for both outsiders and insiders,
encoding for each observer the states hidden behind their illusory horizon.

An infaller who nears the singularity has the impression that
they actually reach the illusory horizon.
This motivates the most speculative proposal in this paper,
that an infaller who hits the singularity merges their states
with the illusory horizon,
the holographic image of the black hole.
In this picture,
the holographic boundary of the black hole
is the union of the illusory horizon with the spacelike singularity.

\section*{Acknowledgements}

I thank Gavin Polhemus for numerous helpful conversations.

\section*{References}
\bibliography{bh}

\begin{thebibliography}{10}

\bibitem{Bousso:2002ju}
Bousso, Raphael, ``The holographic principle'', {\em Rev. Mod. Phys.}, {\bf
  74}, 825--874, (2002).
  {\small[\href{http://dx.doi.org/10.1103/RevModPhys.74.825}{DOI}]},
  {\small[\href{http://arxiv.org/abs/hep-th/0203101}{{arXiv:hep-th/0203101}}]}.

\bibitem{Crispino:2007eb}
Crispino, Luis~C.B., Higuchi, Atsushi  and Matsas, George~E.A., ``The {U}nruh
  effect and its applications'', {\em Rev.Mod.Phys.}, {\bf 80}, 787--838,
  (2008). {\small[\href{http://dx.doi.org/10.1103/RevModPhys.80.787}{DOI}]},
  {\small[\href{http://arxiv.org/abs/0710.5373}{{arXiv:0710.5373
  {\small[gr-qc]}}}]}.

\bibitem{Hamilton:2010my}
Hamilton, Andrew~J.S.  and Polhemus, Gavin, ``Stereoscopic visualization in
  curved spacetime: seeing deep inside a black hole'', {\em New J.\ Phys.},
  {\bf 12}, 123027, (2010).
  {\small[\href{http://dx.doi.org/10.1088/1367-2630/12/12/123027}{DOI}]},
  {\small[\href{http://arxiv.org/abs/1012.4043}{{arXiv:1012.4043
  {\small[gr-qc]}}}]}.

\bibitem{Hamilton:1998coll}
Hamilton, Andrew J.~S., ``Collapse to a black hole'',  (1998)URL:
  \newline\url{http://casa.colorado.edu/~ajsh/collapse.html}.

\bibitem{Hamilton:2010schw}
Hamilton, Andrew J.~S., ``Journey into a {S}chwarzschild black hole'',
  (2010)URL: \newline\url{http://jila.colorado.edu/~ajsh/insidebh/schw.html}.

\bibitem{Hawking:1975b}
Hawking, Stephen~W., ``Particle creation by black holes'', {\em Commun.\ Math.\
  Phys.}, {\bf 43}, 199--220, (1975).

\bibitem{Hawking:1976ra}
Hawking, Stephen~W., ``Breakdown of Predictability in Gravitational Collapse'',
  {\em Phys. Rev.}, {\bf D14}, 2460--2473, (1976).
  {\small[\href{http://dx.doi.org/10.1103/PhysRevD.14.2460}{DOI}]}.

\bibitem{Hodgkinson:2012mr}
Hodgkinson, Lee  and Louko, Jorma, ``Static, stationary and inertial
  {U}nruh-{D}e{W}itt detectors on the {BTZ} black hole'', (2012).
  {\small[\href{http://arxiv.org/abs/1206.2055}{{arXiv:1206.2055
  {\small[gr-qc]}}}]}.

\bibitem{Mellinger:2009asp}
Mellinger, Axel, ``A Color All-Sky Panorama Image of the {M}ilky {W}ay'', {\em
  Pub. Astron. Soc. Pacific}, {\bf 121}, 1180--1187, (2009).

\bibitem{Oppenheimer:1939ue}
Oppenheimer, J.~R.  and Snyder, H., ``On continued gravitational contraction'',
  {\em Phys. Rev.}, {\bf 56}, 455--459, (1939).
  {\small[\href{http://dx.doi.org/10.1103/PhysRev.56.455}{DOI}]}.

\bibitem{Padmanabhan:2010}
Padmanabhan, T., {\em Gravitation: Foundations and Frontiers}, (Cambridge
  University Press, 2010).

\bibitem{Susskind:1994vu}
Susskind, Leonard, ``The {W}orld as a hologram'', {\em J.Math.Phys.}, {\bf 36},
  6377--6396, (1995). {\small[\href{http://dx.doi.org/10.1063/1.531249}{DOI}]},
  {\small[\href{http://arxiv.org/abs/hep-th/9409089}{{arXiv:hep-th/9409089
  {\small[hep-th]}}}]}.

\bibitem{'tHooft:1993gx}
't~Hooft, Gerard, ``Dimensional reduction in quantum gravity'', (1993).
  {\small[\href{http://arxiv.org/abs/gr-qc/9310026}{{arXiv:gr-qc/9310026
  {\small[gr-qc]}}}]}.

\bibitem{Unruh:1976db}
Unruh, W.~G., ``Notes on black hole evaporation'', {\em Phys. Rev.}, {\bf D14},
  870, (1976). {\small[\href{http://dx.doi.org/10.1103/PhysRevD.14.870}{DOI}]}.

\bibitem{Visser:2001kq}
Visser, Matt, ``Essential and inessential features of Hawking radiation'', {\em
  Int. J. Mod. Phys.}, {\bf D12}, 649--661, (2003).
  {\small[\href{http://dx.doi.org/10.1142/S0218271803003190}{DOI}]},
  {\small[\href{http://arxiv.org/abs/hep-th/0106111}{{arXiv:hep-th/0106111
  {\small[hep-th]}}}]}.

\bibitem{Wald:1999vt}
Wald, Robert~M., ``The thermodynamics of black holes'', {\em Living Rev. Rel.},
  {\bf 4}, 6, (2001).
  {\small[\href{http://arxiv.org/abs/gr-qc/9912119}{{arXiv:gr-qc/9912119}}]}.

\bibitem{Wallace:2008zz}
Wallace, Colin~S., Hamilton, Andrew J.~S.  and Polhemus, Gavin, ``Huge entropy
  production inside black holes'', {\em arXiv}, e-print, (2008).
  {\small[\href{http://arxiv.org/abs/0801.4415}{{arXiv:0801.4415
  {\small[gr-qc]}}}]}.

\end{thebibliography}

\end{document}